\documentclass[a4paper,11pt]{article}
\usepackage{jheppub} 
\usepackage{float}
\usepackage[numbers,sort&compress]{natbib}
\usepackage[all]{xy}
\usepackage{amsmath,amsfonts,amssymb,amsthm,epsfig,amscd,comment,latexsym,psfrag,textcomp}
\usepackage{CJK}
\usepackage{mathrsfs}
\usepackage{nicefrac,xspace,tikz}
\usepackage{arydshln}
\usetikzlibrary{arrows,calc}

\usepackage{graphicx}
\usepackage{xcolor,float}
\usepackage{tikz}
\usepackage{float}
\usepackage{graphicx}
\usetikzlibrary{calc}
\usepackage{appendix}

\usepackage{tikz}
\usetikzlibrary{calc}
\usepackage{xstring}
\usepackage{amsmath}
\usepackage{graphicx}
\usepackage{float}
\usepackage{graphicx}
\usepackage{tikz}

\usetikzlibrary{arrows}
\makeatletter

\newcommand{\Rmnum}[1]{\expandafter\@slowromancap\romannumeral #1@}
\makeatother
\usepackage{tikz}

\def\footnoterule{\kern 1mm \hrule width 7cm \kern 2.2mm}%

\allowdisplaybreaks

\usepackage{lineno}

\title{\boldmath \Large\bf Relations between the higher Hamiltonians of the trigonometric and the rational spin Calogero-Sutherland models}

\author[a]{Yue Li,}
\author[b]{Fan Liu,}
\author[c,d]{Lu-Yao Wang}
\author[a,*]{and Jie Yang\note[*]{Corresponding author.}}

\affiliation[a]{School of Mathematical Sciences, Capital Normal University,\\
Beijing 100048, China}

\affiliation[b]{Beijing International Center for Mathematical Research,\\
Beijing 100871, China}

\affiliation[c]{Beijing Institute of Mathematical Sciences and Applications,\\
Beijing 101408, China}

\affiliation[d]{Mathematical Sciences Center, Tsinghua University,\\
Beijing 100084, China}

\begin{document}
\emailAdd{liyue$_{-}$math@cnu.edu.cn}
\emailAdd{liufan$_{-}$math@pku.edu.cn}
\emailAdd{wangluyao@bimsa.cn}
\emailAdd{yangjie@cnu.edu.cn}

\abstract{
In this paper we study the relations between the Hamiltonian hierarchy generated by trigonometric Cherednik-Dunkl operators and the one generated
by rational Dunkl operators.
We develop two nested structures relating these two hierarchies. The first nested relation reconstructs the higher rational spin Calogero-Sutherland
Hamiltonians exactly from the trigonometric ones.
The second nested relation reconstructs the higher trigonometric spin Calogero-Sutherland
Hamiltonians as leading terms from the rational ones.
In addition, we provide an explanation of the eigenvalues of  the trigonometric spin Calogero-Sutherland  Hamiltonians  in terms of
 $N$-colored Young diagrams and Maya diagrams.}

\maketitle
\raggedbottom

\section{Introduction}
The scalar Calogero-Sutherland (CS) models are
one-dimensional quantum integrable many-body systems with long-range
interactions. The rational CS model describes particles on the
line interacting through inverse-square potentials
\cite{Calogero:1969ie,Calogero:1970nt}, while the
trigonometric CS model describes particles on a circle with
periodic inverse-square-type interactions
\cite{PhysRevA.4.2019,PhysRevA.5.1372}.
After a ground-state gauge transformation, the wave functions of these models can be described in terms of special families of  polynomials: Jack polynomials for the trigonometric CS model \cite{STANLEY198976}, and generalized Hermite polynomials for the rational CS model with harmonic confinement \cite{Baker:1996kq,Feigin_2021}.
Both models provide fundamental examples in quantum
many-body integrability and symmetric functions
\cite{STANLEY198976, Baker:1996kq}.

The rational and trigonometric CS models are related in several aspects. Firstly, from the potential aspect, the trigonometric inverse-square interaction can reduce to the rational one by taking a limit of certain parameter \cite{Pasquier:1994cs}.
Secondly, from the algebraic aspect, the rational Hamiltonian
hierarchies are formulated in terms of rational Dunkl operators \cite{Dunkl:1989dvp,CHALYKH2025309}, whereas the
trigonometric hierarchies are formulated in terms of trigonometric
Cherednik-Dunkl operators \cite{Cherednik1991AUO,CHALYKH2025309}. 
These Hamiltonians are conserved charges of quantum integrable systems.
In this way these two models share closely related integrable structures. Thirdly, from the spectral aspect, the two models are related through
their eigenfunctions. Lassalle \cite{Lassalle1991PolynmesDH,Baker:1996kq} constructed multivariable generalized
Hermite polynomials from Jack polynomials, while Nekrasov \cite{Nekrasov:1997jf} established
a classical and quantum correspondence between the trigonometric CS
system and the rational CS system with harmonic confinement. 
The (higher) Hamiltonians and quasi invariant extensions of this correspondence were established in Refs. \cite{Baker:1996kq,Feigin_2021}.

Among these aspects, integrability  gives rise to a commuting hierarchy of the second and higher level of Hamiltonians. It is natural
to ask whether the rational and trigonometric systems can be related
directly at the level of their commuting hierarchies.
In the scalar CS model, the higher trigonometric Hamiltonians were recently constructed in the
Maulik-Okounkov \(R\)-matrix framework \cite{maulik2018}, and suitable nested commutator
structures were shown to reproduce the rational higher Hamiltonians
\cite{2026YLi}.

The spin Calogero-Sutherland (sCS) models extend the scalar CS
models by incorporating internal spin degrees of freedom and
spin-exchange interactions
\cite{1992Haldane}.
The exchange-operator formalism for
many-body systems was developed by Polychronakos
\cite{1992Polychronakos}. Minahan and Polychronakos \cite{MINAHAN1993265} further studied more integrable systems with internal
degrees of freedom. 
Around the same time, the integrability of the sCS model
was established by Hikami and Wadati \cite{Hikami1993}. 
At the operator level, the gauge-transformed sCS Hamiltonians
and their commuting conserved quantities admit realizations
in terms of Dunkl-type operators
\cite{Bernard:1993va,Bernard:1994eb,
	Uglov:1997ia,Takemura:1996qv,Takemura:1997su}.
For the trigonometric sCS model, an algebraic connection with long-range quantum spin chains is provided by Yangian symmetry
\cite{Bernard:1993va,Bernard:1994eb}.
This symmetry accounts for spectral degeneracies and underlies
the construction of eigenbases through Yangian representation
theory \cite{Bernard:1993va,Uglov:1997ia}.

For the finite-particle trigonometric sCS model, Uglov
\cite{Uglov:1997ia} constructed Yangian Gelfand-Zetlin bases
and \(\mathfrak{gl}_N\)-Jack eigenfunctions.
 Takemura and Uglov
\cite{Takemura:1996qv} established an orthogonal eigenbasis
and the corresponding norm formulas.
For the finite-particle rational sCS model with harmonic
confinement, Takemura \cite{Takemura:1997su} used Yangian
representation theory to decompose the state space into
irreducible modules and to construct an orthogonal eigenbasis.
In the infinite-particle limit, Khoroshkin et al.
\cite{Khoroshkin_2017} constructed a multicomponent
Fock-space realization of the trigonometric sCS model
in the infinite-particle limit and obtained explicit formulas
for its Dunkl operators, Yangian generators, and commuting
Hamiltonians. These results show that Yangian representation theory
and Dunkl-type operators provide a common algebraic framework for both
rational and trigonometric sCS models, thereby motivating a direct
comparison of their higher Hamiltonian hierarchies
\cite{Uglov:1997ia,Takemura:1997su,Takemura:1996qv,
Khoroshkin_2017}.

 In this paper, we investigate algebraic relations linking the trigonometric and rational sCS Hamiltonian hierarchies in both directions. We establish two nested-commutator
relations between the higher Hamiltonians generated by trigonometric
Cherednik-Dunkl operators and those generated by rational Dunkl operators.
We also discuss the representation of the eigenbasis.  Based on Uglov's wedge-space construction, we establish an explicit correspondence between the wedge labels and the relating \(N\)-colored Young diagrams as well as Maya diagrams.

The paper is organized as follows. In Section \ref{eq:gauge sCS}, we review the
gauge-transformed rational and trigonometric sCS Hamiltonians, the corresponding Dunkl-type operator
hierarchies, and the antisymmetric coordinate-spin space on which both types of Dunkl operators can act.
 In Section \ref{sec:tritorat}, we introduce a  nested relation, which reconstructs the higher rational sCS
Hamiltonians exactly from the trigonometric ones.
In Section \ref{sec:rattotri},
we propose a reverse nested relation, which reconstructs the higher trigonometric sCS
Hamiltonians as leading terms from the rational ones. In Section \ref{sec:eigenvalues}, we study the trigonometric sCS Hamiltonian eigenvalues in
terms of \(N\)-colored Young diagrams and Maya diagrams. In Section
\ref{sec:summary}, we summarize the results and discuss possible future work.

\section{Gauge-transformed sCS Hamiltonians and Dunkl-type operators}  \label{eq:gauge sCS}
In this section, we recall the standard gauge transformations and Dunkl realizations of the rational and trigonometric sCS Hamiltonians.
\subsection{Gauge-transformed sCS Hamiltonians }

The rational sCS Hamiltonian \cite{Hikami1993,MINAHAN1993265,Takemura:1997su} can be written as
\begin{equation}\label{rat_sCS}
\mathcal{H}_{\mathrm{spin}}^{\mathrm{rat}}
=
-\frac{1}{2}\sum_{i=1}^{n}
\frac{\partial^2}{\partial z_i^2}
+
\sum_{i<j}
\frac{\beta(\beta+P_{ij})}{(z_i-z_j)^2},
\end{equation}
whereas its trigonometric counterpart  \cite{Bernard:1993va,Uglov:1997ia,Takemura:1996qv} is
\begin{equation}\label{tri_sCS}
\mathcal{H}_{\mathrm{spin}}^{\mathrm{tri}}
=
-\frac{1}{2}\sum_{i=1}^{n}
\frac{\partial^2}{\partial x_i^2}
+
\frac{\pi^2}{2L^2}
\sum_{1\leq i\neq j\leq n}
\frac{\beta(\beta+P_{ij})}
{\sin^2\!\left(\frac{\pi}{L}(x_i-x_j)\right)},
\end{equation}
where \(P_{ij}\) denotes the spin-exchange operator.

For the rational model, following Ref.~\cite{Takemura:1997su},
the Jastrow factor is introduced as
\[
\phi_0
=
\prod_{1\leq j<k\leq n}|z_j-z_k|^\beta .
\]
The corresponding gauge-transformed  Hamiltonian is defined by
\begin{equation}\label{gauge_rat_sCH}
\begin{aligned}
H_{\mathrm{spin}}^{\mathrm{rat}}
=&
-2\phi_0^{-1}
\mathcal{H}_{\mathrm{spin}}^{\mathrm{rat}}
\phi_0 \\
=&
\sum_{j=1}^{n}\frac{\partial^2}{\partial z_j^2}
+
2\beta
\sum_{1\leq j<k\leq n}
\frac{1}{z_j-z_k}
\left(
\frac{\partial}{\partial z_j}
-
\frac{\partial}{\partial z_k}
\right)
-2\beta
\sum_{1\leq j<k\leq n}
\frac{1+P_{jk}}{(z_j-z_k)^2}.
\end{aligned}
\end{equation}

For the trigonometric sCS model, following Refs.~\cite{Bernard:1993va,Uglov:1997ia}, the gauge factor is introduced as
\[
\phi
=
\prod_{1\leq i<j\leq n}
\sin\!\left(\frac{\pi}{L}(x_i-x_j)\right).
\]
The corresponding gauge-transformed  Hamiltonian is defined by
\begin{equation}\label{gauge_tri_sCH}
H_{\mathrm{spin}}^{\mathrm{tri}}
=
\frac{L^2}{2\pi^2}
\phi^{-\beta}
\mathcal{H}_{\mathrm{spin}}^{\mathrm{tri}}
\phi^\beta .
\end{equation}
By introducing the exponential coordinates
\[
z_j=
\exp\!\left(\frac{2\pi\sqrt{-1}}{L}x_j\right),
\]
we obtain \cite{Uglov:1997ia}
\begin{equation}\label{eq:gauge-tri-explicit}
\begin{aligned}
H_{\mathrm{spin}}^{\mathrm{tri}}
={}&
\sum_{i=1}^n
\left(z_i\frac{\partial}{\partial z_i}\right)^2
+
\beta\sum_{i=1}^n
(2i-n-1)z_i\frac{\partial}{\partial z_i}
\\
&+
2\beta
\sum_{1\leq i<j\leq n}
\theta_{ij}
\left(
z_i\frac{\partial}{\partial z_i}
-
z_j\frac{\partial}{\partial z_j}
+
\theta_{ji}(P_{ij}+1)
\right)
+
\frac{\beta^2n(n^2-1)}{12},
\end{aligned}
\end{equation}
where
$
\theta_{ij}=\frac{z_i}{z_i-z_j}.$

The above gauge-transformed Hamiltonians admit an algebraic description in terms of commuting Dunkl-type operators, which provides a uniform construction of the higher conserved Hamiltonians as symmetric polynomials of these commuting operators.

\subsection{Trigonometric and rational Dunkl operators}
The trigonometric Cherednik-Dunkl operators  \cite{Uglov:1997ia}
\begin{equation}\label{eq:tri_di}
\hat d_i
=
\beta^{-1}z_i \frac{\partial}{\partial z_i} - i
+ \sum_{i<j}\theta_{ji}(K_{ij}-1)
- \sum_{i>j}\theta_{ij}(K_{ij}-1)\nonumber
\end{equation}
satisfy the commutation relation
\begin{equation}
[\hat d_i,\hat d_j]=0, \nonumber
\end{equation}
where the coordinate
permutation operator $K_{ij}$ exchanges \(z_i\) and \(z_j\). 

The individual Cherednik-Dunkl operators are not invariant under
coordinate permutations. However, symmetric polynomials in these
operators are permutation invariant. This standard property follows
from the defining relations of the degenerate affine Hecke algebra
\cite{Cherednik1991AUO,CHALYKH2025309}. In particular, the power-sum
operators
\[
P_m:=\sum_{a=1}^{n}\hat d_a^m,
\qquad m\geq1,
\]
satisfy the relation
\[
[P_m,K_{ij}]=0,
\qquad
1\leq i<j\leq n .
\]
Moreover, every symmetric polynomial in
\(\hat d_1,\ldots,\hat d_n\) commutes with the coordinate permutation
operators \(K_{ij}\).

It is convenient to introduce the shifted and rescaled Cherednik-Dunkl operators \cite{Uglov:1997ia}
\begin{equation}\label{tri di}
d_i:=\beta\left(\hat d_i+\frac{n+1}{2}\right).
\end{equation}
 Since the operators $d_i$ commute with each other, the commuting conserved quantities can be defined as
\begin{equation}\label{eq:trig-hamiltonian}
\hat H_m^{\mathrm{tri}}
:=
\sum_{i=1}^{n}d_i^m,
\qquad m\geq1.
\end{equation}
For $m=2$, a direct computation gives rise to
\begin{equation}\label{eq:trig-dunkl-square}
\begin{aligned}
\sum_{i=1}^n d_i^2
={}&
\sum_{i=1}^n
\left(z_i\frac{\partial}{\partial z_i}\right)^2
+
\beta\sum_{i=1}^n(2i-n-1)z_i\frac{\partial}{\partial z_i}
\\
&+
2\beta\sum_{1\leq i<j\leq n}
\theta_{ij}
\left(
z_i\frac{\partial}{\partial z_i}
-
z_j\frac{\partial}{\partial z_j}
-
\theta_{ji}(K_{ij}-1)
\right)
+
\frac{\beta^2n(n^2-1)}{12}.
\end{aligned}
\end{equation}

The rational Dunkl operators
\cite{Takemura:1997su,sergeev2013}
\begin{equation}\label{rDunkl}
\mathcal D_i
:=
\frac{\partial}{\partial z_i}
+
\beta\sum_{j\neq i}
\frac{1-K_{ij}}{z_i-z_j},
\end{equation}
satisfy the commutation relation
\begin{equation}\label{eq:nest1}
[\mathcal D_i,\mathcal D_j]=0.
\end{equation}
Similarly,
based on the rational Dunkl operators, the commuting conserved quantities are defined as
\begin{equation}\label{eq:rational-hamiltonian}
\hat H_m^{\mathrm{rat}}
:=
\sum_{i=1}^{n}\mathcal D_i^m,
\qquad m\geq1.
\end{equation}
For \(m=2\), a direct calculation yields
\begin{equation}\label{eq:rational-dunkl-square}
\begin{aligned}
\sum_{i=1}^n\mathcal D_i^2
={}&
\sum_{i=1}^{n}(\frac{\partial}{\partial z_i})^2
+
2\beta\sum_{1\leq i<j\leq n}
(\frac{\partial}{\partial z_i}-\frac{\partial}{\partial z_j} )\frac{1}{z_i-z_j}
\\
&-
2\beta\sum_{1\leq i<j\leq n}
\frac{1-K_{ij}}{(z_i-z_j)^2}.
\end{aligned}
\end{equation}

The sCS models are obtained by extending the coordinate space to an antisymmetric coordinate-spin space $F_{N,n}$, in which the spin degrees of freedom are encoded.  Following
Uglov's construction \cite{Uglov:1997ia}, we define the spin space
\(V=\mathbb C^N\) and work on the antisymmetric coordinate-spin space
\begin{equation}\label{eq:antisymmetric-space}
F_{N,n}
=
\bigwedge^n\!\left(\mathbb C[z^{\pm1}]\otimes V\right)
\subset
\mathbb C[z_1^{\pm1},\ldots,z_n^{\pm1}]
\otimes V^{\otimes n}.
\end{equation}
On this space, the simultaneous exchange of a coordinate variable
and spin variable is antisymmetric. Namely,
the coordinate and spin exchange operators satisfy the relation
\begin{equation}\label{eq:K-P-relation}
K_{ij}f=-P_{ij}f,
\qquad
f\in F_{N,n}.
\end{equation}
After restricting to the antisymmetric coordinate-spin space
$F_{N,n}$, the coordinate-exchange operators \(K_{ij}\)
appearing in the Dunkl-type operators are replaced by the
physical spin-exchange operators \(-P_{ij}\).
Then each symmetric combinations of Dunkl-type operators become the
operators acting on the physical sCS space.
In particular, the operators 
\[H_m^{\mathrm{tri}}:= 
\hat H_m^{\mathrm{tri}}|_{F_{N,n}},
\qquad
H_m^{\mathrm{rat}}:=\hat H_m^{\mathrm{rat}}|_{F_{N,n}}
\]
give the higher Hamiltonians of the trigonometric and rational sCS
models, respectively.
For $m=2$ in the trigonometric case, we have
\begin{equation}\label{eq:tri-second-hamiltonian}
\left.\sum_{i=1}^n d_i^2\right|_{F_{N,n}}
=
H_{\mathrm{spin}}^{\mathrm{tri}}.
\end{equation}
Similarly, for $m=2$ in the rational case, we obtain
\begin{equation}\label{eq:rat-second-hamiltonian}
\left.\sum_{i=1}^n\mathcal D_i^2\right|_{F_{N,n}}
=
H_{\mathrm{spin}}^{\mathrm{rat}}.
\end{equation}

\section{Nested relations between the trigonometric and rational higher Hamiltonians}
\label{sec:nested-relations}

In this section, we establish two nested-commutator relations between the
trigonometric and rational sCS Hamiltonian hierarchies. The first nested relation reconstructs the higher rational sCS
Hamiltonians exactly from the trigonometric ones.
The second nested relation reconstructs the higher trigonometric sCS
Hamiltonians as leading terms from the rational ones.

\subsection{From the trigonometric higher Hamiltonians to the rational higher Hamiltonians}
\label{sec:tritorat}

In this subsection, we show that the higher trigonometric  commuting conserved quantities can be transformed into the corresponding rational commuting conserved quantities
through repeated adjoint actions of the ladder operator
\(E_{-1}\).

We first introduce the ladder operator
\begin{equation}
E_{-1}
=
\sum_{i=1}^{n}\frac{\partial}{\partial z_i}
.
\end{equation}
A direct computation gives the basic commutation relations
\begin{equation}
[E_{-1},d_i]=\mathcal D_i,
\qquad
[E_{-1},\mathcal D_i]=0 .
\end{equation}
These relations show that the adjoint action of \(E_{-1}\) replaces
each trigonometric Cherednik-Dunkl operator \(d_i\) by the rational
Dunkl operator \(\mathcal D_i\). 
It is easy to calculate that
\begin{equation}
\operatorname{ad}_{E_{-1}}^{\,2}(d_i)=0,\nonumber
\end{equation}
where 
$
\operatorname{ad}_{E_{-1}}(X):=[E_{-1},X].
$
Hence the exponential of the adjoint action truncates to
\begin{equation}
e^{t\,\operatorname{ad}_{E_{-1}}}(d_i)
=
d_i+t[E_{-1},d_i]
=
d_i+t\,{\mathcal D_i}.\nonumber
\end{equation}
Since \(e^{t\,\operatorname{ad}_{E_{-1}}}\) is an algebra automorphism, we obtain
\begin{equation} \label{t coeffi}
e^{t\,\operatorname{ad}_{E_{-1}}}(d_i^{\,m})
=
\bigl(e^{t\,\operatorname{ad}_{E_{-1}}}(d_i)\bigr)^m
=
\bigl(d_i+t\,{\mathcal D_i}\bigr)^m.
\end{equation}
On the other hand, expanding the exponential adjoint action in powers of \(t\), we have
\begin{equation}\label{eq:exp-ad-di-m}
e^{t\,\operatorname{ad}_{E_{-1}}}(d_i^{\,m})
=
\sum_{r=0}^{m}\frac{t^r}{r!}\operatorname{ad}_{E_{-1}}^{\,r}(d_i^{\,m}).
\end{equation}
Comparing the coefficients of \(t^m\) in (\ref{eq:exp-ad-di-m}) and
(\ref{t coeffi}),
we observe that they satisfy the following relation
\begin{equation}
\frac{1}{m!}\operatorname{ad}_{E_{-1}}^{\,m}(d_i^m)
=
\mathcal D_i^m .
\end{equation}
Summing over \(i\), we obtain the following nested structure
\begin{equation}\label{eq:tri_to_rat}
\operatorname{ad}_{E_{-1}}^{\,m}\!\left(\sum_{i=1}^n d_i^{\,m}\right)
=
\,m!\sum_{i=1}^n {\mathcal D_i}^{m}.
\end{equation}

Since the above nested relation (\ref{eq:tri_to_rat}) is an identity between Dunkl-type operators,
it remains valid after restricting the operators to the antisymmetric
coordinate-spin space \(F_{N,n}\). Therefore it gives the corresponding
nested relation between the higher Hamiltonians of the trigonometric and
rational sCS models.
When $N=1$, \(K_{ij}\) in the relation \eqref{eq:tri_to_rat} acts as the identity on symmetric functions. Thus the nested structure in (\ref{eq:tri_to_rat}) reduces to the scalar case studied in \cite{2026YLi}.

\subsection{From the rational higher Hamiltonians to the trigonometric higher Hamiltonians}
\label{sec:rattotri}

In this section, we establish the reverse nested structure between the rational
and trigonometric sCS Hamiltonian hierarchies.
Unlike the nested structure in
 section \ref{sec:tritorat}, 
the reverse construction involves correction terms. 

We first introduce the operator
\begin{equation}\label{eq:r2t-E1}
E_1=\sum_{i=1}^{n}z_i^2\frac{\partial}{\partial z_i}
\end{equation}
and compute the following commutation relations
\begin{align*}
[E_1,K_{ij}]&=0,
&[E_1,\mathcal D_i]
&=-2z_i\mathcal D_i+\beta\sum_{j\neq i}(1-K_{ij}),\\
\operatorname{ad}_{E_1}^{\,2}(\mathcal D_i)
&=2z_i^2\mathcal D_i-2\beta z_i\sum_{j\neq i}(1-K_{ij}),
&\operatorname{ad}_{E_1}^{\,3}(\mathcal D_i)&=0.
\end{align*}
Therefore, the exponential adjoint action truncates as
\begin{align*}
e^{t\operatorname{ad}_{E_1}}\mathcal D_i
&=
\mathcal D_i
+t[E_1,\mathcal D_i]
+\frac{t^2}{2}[E_1,[E_1,\mathcal D_i]]
\\
&=
(1-tz_i)^2\mathcal D_i
+
\beta t(1-tz_i)
\sum_{j\neq i}(1-K_{ij}).
\end{align*}
Since \(e^{t\operatorname{ad}_{E_1}}\) is an algebra automorphism, we have
\begin{align}\label{t coeff}
e^{t\operatorname{ad}_{E_1}}(\mathcal D_i^m)
=&
\left(
e^{t\operatorname{ad}_{E_1}}\mathcal D_i
\right)^m  \nonumber \\
=&\left[
(1-tz_i)^2\mathcal D_i
+
\beta t(1-tz_i)
\sum_{j\neq i}(1-K_{ij})
\right]^m \nonumber \\
=&[(1-tz_i)^{2}\mathcal D_i]^m \nonumber\\
&+\beta t\sum_{r=0}^{m-1}\sum_{j\neq i}
\bigl[(1-tz_i)^2\mathcal D_i\bigr]^r
(1-tz_i)(1-K_{ij})
\bigl[(1-tz_i)^2\mathcal D_i\bigr]^{m-1-r}
+\cdots\nonumber  \\
=&(1-tz_i)^{2m}\mathcal D_i^m+\cdots .
\end{align}

where the dots denote those terms whose $\mathcal D_i$ powers are strictly lower
than $m$.

Expanding the exponential adjoint action in powers of \(t\), we have
\begin{equation}\label{eq:exp-ad-Di-m}
e^{t\operatorname{ad}_{E_1}}(\mathcal D_i^m)
=
\sum_{r\geq0}
\frac{t^r}{r!}
\operatorname{ad}_{E_1}^{r}(\mathcal D_i^m).
\end{equation}
Comparing the coefficients of \(t^m\) in (\ref{t coeff}) and
(\ref{eq:exp-ad-Di-m}), we obtain
\begin{align}
   \operatorname{ad}_{E_1}^{m}(\mathcal D_i^m)
=&
(-1)^m
\frac{(2m)!}{m!}
z_i^m\mathcal D_i^m+\cdots\nonumber \\
=&(-1)^m
\frac{(2m)!}{m!}
(z_i\mathcal D_i)^m+\cdots,
\end{align}
where the lower $\mathcal D_i$ powers terms have been omitted.

Since \(z_i\mathcal D_i\) are related to the shifted trigonometric
Cherednik-Dunkl operator \(d_i\) as follows
\begin{eqnarray}
z_i\mathcal D_i
&=&
z_i\frac{\partial}{\partial z_i}
+
\beta\sum_{j\neq i}
\frac{z_i}{z_i-z_j}(1-K_{ij})
\nonumber\\
&=&
d_i
+
\beta
\left(
\frac{n-1}{2}
-
\sum_{j>i}K_{ij}
\right).
\end{eqnarray}
We obtain
\[
\operatorname{ad}_{E_1}^{m}(\mathcal D_i^m)
=
(-1)^m
\frac{(2m)!}{m!}
d_i^m+\cdots .
\]
Summing over \(i\), we get the reverse nested relation
\begin{equation}\label{rat to tri}
\operatorname{ad}_{E_1}^{m}
\left(
\sum_{i=1}^{n}\mathcal D_i^m
\right)
=
(-1)^m
\frac{(2m)!}{m!}
\sum_{i=1}^{n}d_i^m+\cdots
\end{equation}
where the lower powers of $\mathcal D_i$ terms have been omitted like before.

Like section \ref{sec:tritorat}, when restricting the operators on both
sides of \eqref{rat to tri} to \(F_{N,n}\), the rational sCS hierarchy
 reconstructs the trigonometric sCS
hierarchy up to some lower powers of Dunkl operators terms.

When spin $N=1$, the sCS model becomes the scalar CS model. \(K_{ij}\) in the relation \eqref{rat to tri}  acts as the identity on symmetric functions, the nested structure from the rational CS model higher Hamiltonians to the trigonometric ones becomes exact.
For example, the second and third rational CS Hamiltonians \cite{Mironov:2023mve} are taken to be 
\begin{equation}
H_2^{\rm rat}
=
\sum_{i=1}^n (\frac{\partial}{\partial z_i})^2
+
2\beta\sum_{i\neq j}^n
\frac{1}{z_i-z_j}\frac{\partial}{\partial z_i} ,
\end{equation}
and
\begin{equation}
H_3^{\rm rat}
=
\sum_{i=1}^n (\frac{\partial}{\partial z_i})^3
+
3\beta\sum_{ i\neq j}^n
\frac{1}{z_i-z_j}(\frac{\partial}{\partial z_i})^2
+
3\beta^2
\sum_{i\neq j, i\neq k, k\neq j}^n
\frac{1}{(z_i-z_j)(z_i-z_k)}\frac{\partial}{\partial z_i} .
\end{equation}

For the second Hamiltonian, the nested structure gives rise to

\begin{align}\label{H2degenerate: rat to tri}
\operatorname{ad}_{E_1}^2(H_2^{\rm rat})
&=
12\sum_{i=1}^n z_i^2(\frac{\partial}{\partial z_i})^2
+
12\sum_{i=1}^n z_i\frac{\partial}{\partial z_i}
+
12\beta
\sum_{\substack{i,j=1\\ i\neq j}}^n
\frac{z_i(z_i+z_j)}{z_i-z_j}\frac{\partial}{\partial z_i}\nonumber \\
&=12\left(
\sum_{i=1}^n (z_i\frac{\partial}{\partial z_i})^2
+
\beta\sum_{1\leq i<j\leq n}
\frac{z_i+z_j}{z_i-z_j}
(z_i\frac{\partial}{\partial z_i}-z_j\frac{\partial}{\partial z_j})\right) .
\end{align}
For the third Hamiltonian, we have
\begin{equation}\label{degenerate: rat to tri}
\begin{aligned}
\operatorname{ad}_{E_1}^{3}\!\left(H^{\mathrm{rat}}_{3}\right)
={}&
-120\sum_{i=1}^n z_i^3(\frac{\partial}{\partial z_i})^3
-360\sum_{i=1}^n z_i^2(\frac{\partial}{\partial z_i})^2
-144\sum_{i=1}^n z_i\frac{\partial}{\partial z_i}
\\[1mm]
&-180\beta
\sum_{\substack{i,j=1\\ i\neq j}}^n
\frac{z_i^2(z_i+z_j)}{z_i-z_j}
(\frac{\partial}{\partial z_i})^2
-72\beta
\sum_{\substack{i,j=1\\ i\neq j}}^n
\frac{z_i(3z_i+2z_j)}{z_i-z_j}
\frac{\partial}{\partial z_i}
\\[1mm]
&-36\beta^2
\sum_{i\neq j, i\neq k,k\neq j}^n
\frac{
z_i\left(
2z_i^2+3z_iz_j+3z_iz_k+2z_jz_k
\right)
}
{(z_i-z_j)(z_i-z_k)}
\frac{\partial}{\partial z_i}.
\end{aligned}
\end{equation}

In terms of the power-sum variables $
p_m=\sum_{i=1}^{n}z_i^m,
p_0=n$, and $\partial_m:=\frac{\partial}{\partial p_m},
$
 the above nested commutators can be reformulated in terms of symmetric
functions. A double commutator with \(E_1\) gives rise to
\begin{eqnarray}\label{eq:powersum_H2}
\operatorname{ad}_{E_1}^2(H_2^{\rm rat})
&=&12\sum_{a,b>0}\big [\beta(a+b)p_{a}p_{b}\frac{\partial}{\partial p_{a+b}}+abp_{a+b}\frac{\partial}{\partial p_{a}}\frac{\partial}{\partial p_{b}}\big ]
+12\sum_{a>0}(1-\beta)a^{2}p_{a}\frac{\partial}{\partial p_{a}} \nonumber\\
&&+12\beta n\sum_{a>0}ap_{a}\frac{\partial}{\partial p_{a}} .
\end{eqnarray}
Similarly, a direct computation then gives the corresponding triple commutator
\begin{equation}\label{eq:powersum_H3}
\begin{aligned}
\operatorname{ad}_{E_1}^{3}\!\left(H^{\mathrm{rat}}_{3}\right)
={}&-120\sum_{a,b,c\ge 1}abc\,p_{a+b+c}\partial_a\partial_b\partial_c  -120\beta^2\sum_{r,s,t\ge 1}
(r+s+t)p_rp_sp_t\,\partial_{r+s+t} \\
&-180\sum_{a,b\ge 1}ab
\biggl[
\beta\sum_{\substack{r,s\ge 1\\ r+s=a+b}}p_rp_s
+\bigl((1-\beta)(a+b)+\beta n\bigr)p_{a+b}
\biggr]\partial_a\partial_b \\
&-180\sum_{r,s\ge 1}
(r+s)\bigl[\beta(r+s)+\beta^2(n-r-s)\bigr]
p_rp_s\,\partial_{r+s} \\
&-12\sum_{a\ge 1}a
\Bigl[
2(5a^2+1)
+3\beta\bigl((5a+1)n-5a^2-1\bigr) \\
&
+\beta^2\bigl(10a^2-15an+6n^2-3n+2\bigr)
\Bigr]p_a\partial_a .
\end{aligned}
\end{equation}
The detailed derivations of \eqref{eq:powersum_H2} and \eqref{eq:powersum_H3} are presented in the Appendix.  These expressions define (higher) conserved Hamiltonians of the
trigonometric CS hierarchy. After an appropriate normalization, the results of $\operatorname{ad}_{E_1}^{2}\!\left(H^{\mathrm{rat}}_{2}\right) $ and $\operatorname{ad}_{E_1}^{3}\!\left(H^{\mathrm{rat}}_{3}\right)$ reduce
to the canonically normalized second and third Hamiltonians of
Ref.~\cite{2026YLi}.

\section{Eigenvalues of the trigonometric sCS Hamiltonians
labelled by $N$-colored Young diagrams and Maya diagrams}\label{sec:eigenvalues}
In this section, we describe the spectral data of the trigonometric sCS
Hamiltonians in terms of \(N\)-colored Young diagrams and Maya diagrams.
We first recall the Yangian Gelfand-Zetlin basis and the corresponding
eigenvalues of the Cherednik-Dunkl operators. Then we reinterpret the
wedge labels of this basis in terms of these combinatorial objects.

\subsection{Yangian symmetry and the Gelfand-Zetlin basis}

Following Uglov \cite{Uglov:1997ia}, we define a vector space
\[
V(z)=\mathbb{C}[z^{\pm1}]\otimes V,
\qquad
V=\mathbb{C}^N,
\]
whose one-particle basis is
\[
u_k=z^{\overline{k}}\otimes v_{\underline{k}},
\]
where
\begin{equation}
k=\underline{k}-N\overline{k},
\qquad
\underline{k}\in\{1,\ldots,N\},
\qquad
\overline{k}\in\mathbb{Z}.
\end{equation}
The antisymmetric \(n\)-particle coordinate-spin space is
\[
F_{N,n}
\subset
\mathbb{C}[z_1^{\pm1},\ldots,z_n^{\pm1}]
\otimes V^{\otimes n},
\]
with normally ordered wedge basis
\[
\hat{u}_{\mathbf{k}}
=
u_{k_1}\wedge\cdots\wedge u_{k_n},
\qquad
k_1>\cdots>k_n,
\]
where each wedge label admits the unique decomposition
\begin{equation}\label{eq:decomposition}
    k_i=\underline{k_i}-N\overline{k_i},
\qquad
\underline{k_i}\in\{1,\ldots,N\},
\quad
\overline{k_i}\in\mathbb{Z}.
\end{equation}
Here $\underline{k_i}$ is the spin/color label, whereas
$\overline{k_i}$ is the orbital/energy label.
This decomposition is compatible with the Yangian structure underlying
the trigonometric sCS model.

The trigonometric sCS model possesses a Yangian
\(Y(\mathfrak{gl}_N)\) symmetry. The Lax operator relevant to the
commuting Hamiltonians is \cite{Bernard:1993va,Uglov:1997ia}
\begin{equation}
L_{ab}^{(i)}(u;\beta)
=
\delta_{ab}
+
\frac{E_{ab}^{(i)}}{u+\hat d_i},
\end{equation}
where \(\hat d_i\) is the trigonometric Cherednik-Dunkl operator
introduced in \eqref{eq:tri_di}.
The corresponding monodromy matrix defines an action of
\(Y(\mathfrak{gl}_N)\) on \(F_{N,n}\).

The quantum determinant of this Yangian action is
\[
A_N(u;\beta)
=
\prod_{i=1}^{n}
\frac{u+1+\hat d_i}{u+\hat d_i},
\]
whose expansion generates commuting conserved quantities contained in
the maximal commutative subalgebra
$
A(\mathfrak{gl}_N;\beta).
$
A simultaneous eigenbasis of this subalgebra is the Gelfand-Zetlin basis
\cite{Uglov:1997ia}
\begin{equation}\label{eq:GZ basis}
X_{\mathbf{k}}^{(\beta,N)}
=
\hat u_{\mathbf{k}}
+
\sum_{\mathbf l<\mathbf k}
x_{\mathbf{k}\mathbf l}^{(\beta)}
\hat u_{\mathbf l},
\end{equation}
where \(x_{\mathbf{k}\mathbf l}^{(\beta)}\) are the coefficients of the
triangular change of basis from the normally ordered wedge basis
\(\{\hat u_{\mathbf l}\}\) to the Gelfand-Zetlin basis, and
\(\mathbf l<\mathbf k\) denotes the ordering used in the triangular
decomposition.
The Yangian symmetry accounts for the degeneracies of the Hamiltonian
spectrum, while the triangular form in \eqref{eq:GZ basis} allows the
spectral data to be read from the leading wedge label \(\mathbf k\) \cite{Uglov:1997ia}.

\subsection{Eigenvalue of the Cherednik-Dunkl operator }

For later comparison, we decompose the wedge label for the vacuum vector
\(k_i^{\mathrm{vac}}=M-i+1\) in the same way
\[
k_i^{\mathrm{vac}}
=
\underline{k_i^{\mathrm{vac}}}
-
N\overline{k_i^{\mathrm{vac}}},
\qquad
1\leq \underline{k_i^{\mathrm{vac}}}\leq N,
\qquad
\overline{k_i^{\mathrm{vac}}}\in\mathbb Z.
\]
Here \(\underline{k_i^{\mathrm{vac}}}\) and
\(\overline{k_i^{\mathrm{vac}}}\) denote, respectively, the vacuum
color and orbital components.

The symmetric trigonometric Hamiltonians act diagonally on the
Gelfand-Zetlin basis \cite{Uglov:1997ia,Lamers:2022ioe, Ferrando:2023lrx}
\[
H_m^{\mathrm{tri}}X_{\mathbf{k}}^{(\beta,N)}
=
E_m(\mathbf{k})X_{\mathbf{k}}^{(\beta,N)},
\]
with
\[
E_m(\mathbf{k})
=
\sum_{i=1}^{n}
\left[
\overline{k_i}
+
\beta\left(i-\frac{n+1}{2}\right)
\right]^m.
\]
Although the eigenvectors are labelled by the wedge sequence $\mathbf{k}$, their eigenvalues depend only on the orbital
components $\overline{k_i}$ of $\mathbf{k}$.

If we choose a vacuum state which corresponds to a wedge vector
\begin{equation}\label{vac}
	\mathrm{vac}(M)
	=
	u_M\wedge u_{M-1}\wedge\cdots\wedge u_{M-n+1},
\end{equation}
then for a state which corresponds to a Young diagram
\begin{equation}
        \lambda=(\lambda_1,\ldots,\lambda_n),
        \nonumber
\end{equation}
the corresponding labels for its wedge vector are shifted to
\begin{equation}\label{ki}
        k_i=k_i^{\mathrm{vac}}+\lambda_i=M-i+1+\lambda_i.
\end{equation}

For fixed \(i\), the intermediate labels are
\[
k_{i,j}=k_i^{\rm vac}+j=M-i+1+j,
\qquad 0\leq j\leq \lambda_i .
\]
We decompose each \(k_{i,j}\) as
\[
k_{i,j}
=
\underline{k_{i,j}}
-
N\overline{k_{i,j}},
\]
where
\(\underline{k_{i,j}}\in\{1,\ldots,N\}\) and
\(\overline{k_{i,j}}\in\mathbb{Z}\).
Then we have
\[
k_{i,0}=k_i^{\rm vac},
\qquad
k_{i,\lambda_i}=k_i \qquad \overline{k_{i,0}}=\overline{k_i^{\rm vac}}.
\]
When $j$ increases as long as $1\leq j<kN+i-M$ ($k\in \mathbb{Z} $ such that $1\leq kN+i-M\leq N$ ), we have $\overline{k_{i,j}}=\overline{k_i^{\rm vac}}$. However when $j=kN+i-M$,  we obtain $\overline{k_{i,j}}=\overline{k_i^{\rm vac}}-1$.
Likewise we keep increasing $j$ till $j=\lambda_i$ and we find that at every $j$ such that $M-i+j\equiv0\pmod N$,
the orbital component
decreases by one.
Namely
\[
\overline{k_{i,j}}-\overline{k_{i,j-1}}
=
\begin{cases}
-1, & M-i+j\equiv0\pmod N,\\
0,  & \text{otherwise}.
\end{cases}
\]
For a box  $(i,j)$ in the Young diagram,  we define its color number
\begin{equation}
        \operatorname{col}_{M}(i,j)
        =
        1+\bigl((M-i+j)\bmod N\bigr).
        \label{eq:M-dependent-color}
\end{equation} Then
the condition
$
M-i+j\equiv0\pmod N
$
is equivalent to
$
\operatorname{col}_{M}(i,j)=1.
$
Hence each color-\(1\) box contributes exactly $-1$ to the
orbital component. The
difference  $\overline{k_{i}^{\rm vac}}-\overline{k_{i}}$ is determined
precisely by the number of color-\(1\) boxes in the \(i\)-th row.

Then we introduce 
\begin{equation}
        w_i^{(M,N)}(\lambda)
        :=
        \#\{1\leq j\leq\lambda_i\mid M-i+j\equiv0\pmod N\},
        \label{eq:wi-definition}
\end{equation}
which is the number of color-1 boxes in the $i$-th row. Hence we have
\begin{equation}
        \overline{k_i}
        =
        \overline{k_i^{\mathrm{vac}}}
        -
        w_i^{(M,N)}(\lambda).
        \label{eq:kbar-shift}
\end{equation}
Let us illustrate this with an example.
Using the \(M\)-dependent coloring \eqref{eq:M-dependent-color}
and taking
\[
        N=6,\qquad M=2,\qquad \lambda=(8,6,5,3,2),
\]
we obtain the following colored Young diagram
\[
\begin{array}{c|cccccccc}
i\backslash j & 1&2&3&4&5&6&7&8\\
\hline
1 & 3&4&5&6&{\color{red}1}&2&3&4\\
2 & 2&3&4&5&6&{\color{red}1}&&\\
3 & {\color{red}1}&2&3&4&5&&&\\
4 & 6&{\color{red}1}&2&&&&&\\
5 & 5&6&&&&&&
\end{array}
\]
Table 1: The entries in the box \((i,j)\) show the \(M\)-dependent color
\(\operatorname{col}_{M}(i,j)\). The color-1 boxes are marked as red. They contribute to the eigenvalues.
The row lengths and the corresponding orbital components are
\[
\begin{array}{c|ccccc}
i & 1&2&3&4&5\\
\hline
\lambda_i & 8&6&5&3&2\\
\overline{k_i} & -1&-1&0&0&1
\end{array}
\]

For \(m=2\), we obtain
\begin{equation}
\begin{aligned}
E_{2}(k)-E_{2}(k^{\mathrm{vac}})
&=
\sum_i
\left[
\left(
\overline{k_i^{\mathrm{vac}}}
-w_i^{(M,N)}(\lambda)
\right)^2
-
\left(
\overline{k_i^{\mathrm{vac}}}
\right)^2
\right]
-\beta\sum_i(2i-n-1)w_i^{(M,N)}(\lambda)
\\
&=
\sum_i\left(w_i^{(M,N)}(\lambda)\right)^2
-
\sum_i
\left[
2\overline{k_i^{\mathrm{vac}}}
+\beta(2i-n-1)
\right]
w_i^{(M,N)}(\lambda).
\end{aligned}
\label{eq:energy-shift-relevant-color}
\end{equation}
Thus, the energy shift is determined entirely by the color-\(1\)
boxes in the \(M\)-dependent coloring defined in
\eqref{eq:M-dependent-color}. Upon specializing to \(N=2\) and \(M=n/2+1\), this formula reduces to
Uglov's checkerboard description of the momentum and energy
eigenvalues \cite{Uglov:1997ia}.

\subsection{The wedge basis, \texorpdfstring{$N$}{N}-colored Young diagrams and Maya diagrams}

We now reinterpret the wedge labels in terms of Maya
diagrams.  Recall that the \(n\)-particle basis of the antisymmetric
coordinate-spin space is given by normally ordered wedge vectors \cite{Uglov:1997ia}
\[
\hat u_{\mathbf{k}}
=
u_{k_1}\wedge u_{k_2}\wedge \cdots \wedge u_{k_n},
\qquad
k_1>k_2>\cdots>k_n .
\]
Each \(u_{k_i}\) represents an occupied one-particle state located at some position. Hence $\hat u_{\mathbf{k}}$ means a set of particles located at a set of positions. In the meantime a Maya diagram contains black dots and white dots on a one-dimensional lattice, where a black dot represents an occupied state associated with a wedge label \(k_i\) and a white dot for an
unoccupied point.  The wedge vector \eqref{vac}
plays the role of the vacuum Maya diagram. A general wedge vector \eqref{ki} is obtained from this vacuum by moving some of its occupied points to new positions,
which is precisely the combinatorial content of a Young diagram.
The length $\lambda_i$ of the $i$-th row measures the displacement
of the corresponding occupied wedge position relative to the vacuum. For simplicity in the  colored vacuum configuration, we choose \(M=N\).

Let us first consider the case $N=1$, where the decomposition \eqref{eq:decomposition} reduces to
\[
k=\underline{k}-\overline{k}=1-\overline{k}.
\]
We can define the corresponding lattice points for the Maya diagram
\[
r=\overline{k}+\frac12=-k+\frac32.
\]
Hence the occupied points for the vacuum are
\[
r_i^{\mathrm{vac}}
=
-k_i^{\mathrm{vac}}+\frac32
=
-(1-i+1)+\frac32.
\]
In conclusion, the Maya lattice is
\[
\mathbb Z+\frac12
=
\left\{
\cdots,-\frac52,-\frac32,-\frac12,
\frac12,\frac32,\frac52,\cdots
\right\}
\]
and the vacuum Maya diagram is represented as
\begin{center}
\begin{tikzpicture}
  \node at (-0.8, 0) {$\cdots$};

  \foreach \x in {0,...,5} {
    \ifnum\x<3
      \draw (\x, 0) circle[radius=3pt];
    \else
      \fill (\x, 0) circle[radius=3pt];
    \fi
    \edef\myNum{\the\numexpr -5 + 2*\x \relax}
    \ifnum\myNum<0
      \node[below=8pt] at (\x, 0) {$-\frac{\the\numexpr -\myNum \relax}{2}$};
    \else
      \node[below=8pt] at (\x, 0) {$\frac{\myNum}{2}$};
    \fi
  }
,
  \node at (5.8, 0) {$\cdots$};
\end{tikzpicture}
\end{center}
where black and white dots denote occupied and empty sites
respectively.

We now turn to the $N$-colored Maya diagram.
Each wedge label admits the unique decomposition
\[
k=\underline{k}-N\overline{k},
\qquad
\underline{k}\in\{1,\ldots,N\},
\qquad
\overline{k}\in\mathbb Z,
\]
where $\underline{k}$ is the spin component and
$\overline{k}$ is the orbital component.
To encode these two pieces of information in a  diagram,
we associate the wedge label $k$ with the colored Maya point
\[
(r,a)
=
\left(
\overline{k}+\frac12,
N+1-\underline{k}
\right).
\]
The half-integer coordinate $r$ determines the orbital component,
while the color label $a$ determines the spin component.

For the $i$-th row, the occupied points for the $N$-colored vacuum Maya diagram are
\[
(r_i^{\mathrm{vac}},a_i^{\mathrm{vac}})
=
\left(
\overline{k_i^{\mathrm{vac}}}+\frac12,
N+1-\underline{k_i^{\mathrm{vac}}}
\right),
\]
The $N$-colored vacuum fills all color sectors at every positive
half-integer site and leaves the negative sites empty
\begin{center}
	\begin{tikzpicture}
		\pgfmathsetmacro{\spacing}{1.2}   
		\def\cnt{0}                       
		
		\foreach \i in {0,...,10} {
			\pgfmathsetmacro{\x}{\i * \spacing}
			
			\ifcase\i
			\def\type{0} 
			\or
			\def\type{1} 
			\or
			\def\type{1} 
			\or
			\def\type{0} 
			\or
			\def\type{1} 
			\or
			\def\type{2} 
			\or
			\def\type{2} 
			\or
			\def\type{0} 
			\or
			\def\type{2} 
			\or
			\def\type{2} 
			\or
			\def\type{0} 
			\fi
			
			\ifnum\type=0
			\node at (\x, 0) {$\cdots$};
			\else
			\ifnum\type=1
			\draw (\x, 0) circle[radius=3pt];   
			\else
			\fill (\x, 0) circle[radius=3pt];   
			\fi
			
			\ifcase\cnt
			\def\mylabel{$-\frac{1}{2}^{(1)}$} 
			\or
			\def\mylabel{$-\frac{1}{2}^{(2)}$} 
			\or
			\def\mylabel{$-\frac{1}{2}^{(N)}$} 
			\or
			\def\mylabel{$\frac{1}{2}^{(1)}$}  
			\or
			\def\mylabel{$\frac{1}{2}^{(2)}$}  
			\or
			\def\mylabel{$\frac{1}{2}^{(N)}$}  
			\or
			\def\mylabel{$\frac{3}{2}^{(1)}$}  
			\fi
			\node[below=8pt] at (\x, 0) {\mylabel};
			
			\pgfmathsetmacro{\cnt}{int(\cnt+1)}
			\xdef\cnt{\cnt}
			\fi
		}
	\end{tikzpicture}
\end{center}

The excited wedge label \(k_i\) corresponds to an occupied point
in the \(N\)-colored Maya diagram, with position and color given by
\[
(r_i^{\mathrm{occ}},a_i^{\mathrm{occ}})
=
\left(
\overline{k_i}+\frac12,\,
N+1-\underline{k_i}
\right).
\]
Since \(k_i-k_i^{\mathrm{vac}}=\lambda_i\), we have
\[
\begin{aligned}
\lambda_i
&=
\underline{k_i}-\underline{k_i^{\mathrm{vac}}}
-N(\overline{k_i}-\overline{k_i^{\mathrm{vac}}})
\\
&=
a_i^{\mathrm{vac}}-a_i^{\mathrm{occ}}
-
N(r_i^{\mathrm{occ}}-r_i^{\mathrm{vac}}).
\end{aligned}
\]

Thus, for $N>1$, the row length is encoded by the displacement of the Maya site and the change of its color.
The former determines the orbital component, while the latter
determines the spin component.

Having identified the vacuum and occupied colored Maya points,
we can now express the Young diagram in terms of fermion creation and annihilation operators.
The operator
$
\psi_r^{(a)}
$
creates a fermion of color $a$ at the Maya site $r$, while
$
\psi_{-r}^{*(a)}
$
 annihilates a fermion of color $a$ at the site $r$. They satisfy the anticommutation relation
 $
\left\{
\psi_r^{(a)},\psi_s^{*(b)}
\right\}
=
\delta_{ab}\delta_{r+s,0}.
$
The contribution of the $i$-th row is obtained by removing the vacuum
particle at
\[
(r_i^{\mathrm{vac}},a_i^{\mathrm{vac}})
\]
and creating a particle at
\[
(r_i^{\mathrm{occ}},a_i^{\mathrm{occ}}).
\]
Therefore the corresponding fermionic operator is
\[
\psi_{r_i^{\mathrm{occ}}}^{(a_i^{\mathrm{occ}})}
\psi_{-r_i^{\mathrm{vac}}}^{*(a_i^{\mathrm{vac}})} .
\]
Up to an overall fermionic normal-ordering sign,
the colored Young diagram corresponds to
\[
|\lambda\rangle_{\mathrm{Maya}}
=
\left(
\prod_{i=1}^{\ell(\lambda)}
\psi_{r_i^{\mathrm{occ}}}^{(a_i^{\mathrm{occ}})}
\psi_{-r_i^{\mathrm{vac}}}^{*(a_i^{\mathrm{vac}})}
\right)
|0\rangle .
\]

\begin{figure}[H]
\centering
\begingroup

\newcommand{\ycell}[4]{%
  \filldraw[
    fill=#1,
    draw=white,
    line width=0.55pt
  ] ({#2},{-#3}) rectangle ++(1,-1);
  \node[
    text=white,
    font=\bfseries\small
  ] at ({#2+0.5},{-#3-0.5}) {$#4$};
}

\newcommand{\Lmid}[3]{%
  \coordinate (#1) at ({#2},{-#3-0.5});
}
\newcommand{\Rmid}[3]{%
  \coordinate (#1) at ({#2+1},{-#3-0.5});
}
\newcommand{\Tmid}[3]{%
  \coordinate (#1) at ({#2+0.5},{-#3});
}

\newcommand{\MarkIntersection}[3]{%
  \fill[red] (#1) circle (1.45pt);
  \node[
    red,
    font=\bfseries\small
  ] at ($(#1)+#3$) {$#2$};
}


\def\mayay{-0.05}
\def\mayalabeldy{-0.55}

\newcommand{\MayaPointAt}[3]{%
  \coordinate (maya-#1) at (#1 |- mayaBase);
  \node[dot,fill=#2] at (maya-#1) {};
  \node[anchor=north,font=\scriptsize]
    at ($(maya-#1)+(0,\mayalabeldy)$) {#3};
}

\newcommand{\MayaUnknownAt}[2]{%
  \coordinate (maya-#1) at (#1 |- mayaBase);
  \node[
    dot,
    draw=gray!70,
    fill=gray!15
  ] at (maya-#1) {};
  \node[anchor=north,font=\scriptsize]
    at ($(maya-#1)+(0,\mayalabeldy)$) {#2};
}

\newcommand{\MayaEllipsisBetween}[3]{%
  \path (maya-#1) -- (maya-#2)
    node[midway] {$\cdots$}
    node[midway,anchor=north,font=\scriptsize,
         yshift={\mayalabeldy cm}] {#3};
}

\resizebox{0.8\textwidth}{!}{%
\begin{tikzpicture}[
    x=1cm,
    y=1cm,
    dot/.style={
      circle,
      draw=black,
      minimum size=3.5mm,
      inner sep=0pt
    }
]

\tikzset{
  axis/.style={->,semithick},
  proj/.style={
    red,
    dashed,
    line width=0.45pt
  }
}

\begin{scope}[xshift=-1.30cm,yshift=1.45cm,
              rotate around={135:(0,0)}]

\draw[axis] (0,0) -- (8.2,0);
\draw[axis] (0,0) -- (0,-8.2);

\ycell{blue!70!black}{0}{0}{1}
\ycell{teal!70!black}{1}{0}{2}
\ycell{green!55!black}{2}{0}{3}
\ycell{gray!55}{3}{0}{\ddots}
\ycell{orange!85!black}{4}{0}{N-1}
\ycell{purple!75!black}{5}{0}{N}
\ycell{blue!70!black}{6}{0}{1}

\ycell{purple!75!black}{0}{1}{N}
\ycell{blue!70!black}{1}{1}{1}
\ycell{teal!70!black}{2}{1}{2}
\ycell{gray!55}{3}{1}{\ddots}
\ycell{pink!85!black}{4}{1}{N-2}
\ycell{orange!85!black}{5}{1}{N-1}

\ycell{orange!85!black}{0}{2}{N-1}
\ycell{purple!75!black}{1}{2}{N}
\ycell{blue!70!black}{2}{2}{1}
\ycell{gray!55}{3}{2}{\ddots}
\ycell{cyan!60!black}{4}{2}{N-3}

\ycell{gray!55}{0}{3}{\ddots}
\ycell{gray!55}{1}{3}{\ddots}
\ycell{gray!55}{2}{3}{\ddots}
\ycell{gray!55}{3}{3}{\ddots}
\ycell{gray!55}{4}{3}{\ddots}

\ycell{green!55!black}{0}{4}{3}
\ycell{red!85!black}{1}{4}{4}
\ycell{orange!90!black}{2}{4}{5}

\ycell{teal!70!black}{0}{5}{2}
\ycell{green!55!black}{1}{5}{3}

\ycell{blue!70!black}{0}{6}{1}


\Rmid{pAbase}{6}{0}
\Lmid{pBbase}{5}{0}

\Rmid{pCbase}{0}{6}
\Lmid{pGbase}{0}{6}

\Rmid{pHbase}{0}{0}
\Lmid{pDbase}{0}{0}
\Lmid{pEbase}{0}{1}
\Lmid{pFbase}{0}{2}

\foreach \x in {0,1,2,4,5,6}{
  \Tmid{TopMid\x}{\x}{0}
}
\foreach \y in {0,1,2,4,5,6}{
  \Lmid{LeftMid\y}{0}{\y}
}

%
\Tmid{Bouter}{6}{1}
%
\Tmid{Jouter}{1}{6}
%
\Tmid{Louter}{0}{7}

\end{scope}


\def\Adx{0.00}
\def\Ady{0.00}

\def\Bdx{0.00}
\def\Bdy{1.35}

\def\Cdx{0.00}
\def\Cdy{0.00}

\def\Hdx{0.00}
\def\Hdy{-0.70}

\def\Ddx{0.00}
\def\Ddy{0.00}

\def\Edx{0.00}
\def\Edy{0.00}

\def\Fdx{0.00}
\def\Fdy{0.00}

\def\Gdx{0.00}
\def\Gdy{0.00}

\coordinate (pA) at ($(pAbase)+(\Adx,\Ady)$);
\coordinate (pB) at ($(pBbase)+(\Bdx,\Bdy)$);
\coordinate (pC) at ($(pCbase)+(\Cdx,\Cdy)$);
\coordinate (pH) at ($(pHbase)+(\Hdx,\Hdy)$);
\coordinate (pD) at ($(pDbase)+(\Ddx,\Ddy)$);
\coordinate (pE) at ($(pEbase)+(\Edx,\Edy)$);
\coordinate (pF) at ($(pFbase)+(\Fdx,\Fdy)$);
\coordinate (pG) at ($(pGbase)+(\Gdx,\Gdy)$);


\foreach \x in {0,1,2,4,5,6}{
  \fill[red] (TopMid\x) circle (1.2pt);
}
\foreach \y in {0,1,2,4,5,6}{
  \fill[red] (LeftMid\y) circle (1.2pt);
}


\MarkIntersection{pA}{\mathrm{A}}
{(-0.20,0.32)}


\MarkIntersection{pB}{\mathrm{C}}
{(-0.20,0.25)}

\MarkIntersection{TopMid2}{\mathrm{D}}{(-0.12,-0.34)}

\MarkIntersection{TopMid1}{\mathrm{E}}{(-0.12,-0.24)}

\MarkIntersection{TopMid0}{\mathrm{F}}{(-0.35,-0.08)}

\MarkIntersection{pD}{\mathrm{G}}{(0.18,-0.40)}

\MarkIntersection{pE}{\mathrm{H}}{(0.20,-0.25)}

\MarkIntersection{pF}{\mathrm{I}}{(0.25,-0.10)}


\MarkIntersection{pC}{\mathrm{K}}{(0.10,0.42)}


\MarkIntersection{Bouter}{\mathrm{B}}{(-0.12,0.35)}
\MarkIntersection{Jouter}{\mathrm{J}}{(0.22,0.12)}
\MarkIntersection{Louter}{\mathrm{L}}{(0.22,0.12)}


\coordinate (mayaBase) at (0,\mayay);

\MayaPointAt{pA}{black}
  {$\psi^{(N)}_{-\frac32}$}
\MayaPointAt{TopMid5}{white}
  {$\psi^{(1)}_{-\frac12}$}
\MayaPointAt{pB}{black}
  {$\psi^{(2)}_{-\frac12}$}

\MayaUnknownAt{TopMid2}
  {$\psi^{(N-2)}_{-\frac12}$}
\MayaUnknownAt{TopMid1}
  {$\psi^{(N-1)}_{-\frac12}$}
\MayaUnknownAt{TopMid0}
  {$\psi^{(N)}_{-\frac12}$}

\MayaUnknownAt{pD}
  {$\psi^{(1)}_{\frac12}$}
\MayaUnknownAt{pE}
  {$\psi^{(2)}_{\frac12}$}
\MayaUnknownAt{pF}
  {$\psi^{(3)}_{\frac12}$}

\MayaPointAt{LeftMid4}{white}
  {$\psi^{(N-1)}_{\frac12}$}
\MayaPointAt{pC}{black}
  {$\psi^{(N)}_{\frac12}$}
\MayaPointAt{LeftMid6}{white}
  {$\psi^{(1)}_{\frac32}$}

\node at ($(maya-pA)+(-0.85,0)$) {$\cdots$};
\node[anchor=north,font=\scriptsize]
  at ($(maya-pA)+(-0.85,\mayalabeldy)$) {$\cdots$};

\MayaEllipsisBetween{pB}{TopMid2}{$\cdots$}
\MayaEllipsisBetween{pF}{LeftMid4}{$\cdots$}

\node at ($(maya-LeftMid6)+(0.85,0)$) {$\cdots$};
\node[anchor=north,font=\scriptsize]
  at ($(maya-LeftMid6)+(0.85,\mayalabeldy)$) {$\cdots$};


\draw[proj,shorten >=1.75mm] (pA)       -- (maya-pA);
\draw[proj,shorten >=1.75mm] (TopMid5)  -- (maya-TopMid5);
\draw[proj,shorten >=1.75mm] (pB)       -- (maya-pB);
\draw[proj,shorten >=1.75mm] (TopMid2)  -- (maya-TopMid2);
\draw[proj,shorten >=1.75mm] (TopMid1)  -- (maya-TopMid1);
\draw[proj,shorten >=1.75mm] (TopMid0)  -- (maya-TopMid0);
\draw[proj,shorten >=1.75mm] (pD)       -- (maya-pD);
\draw[proj,shorten >=1.75mm] (pE)       -- (maya-pE);
\draw[proj,shorten >=1.75mm] (pF)       -- (maya-pF);
\draw[proj,shorten >=1.75mm] (LeftMid4) -- (maya-LeftMid4);
\draw[proj,shorten >=1.75mm] (pC)       -- (maya-pC);
\draw[proj,shorten >=1.75mm] (LeftMid6) -- (maya-LeftMid6);

\draw[proj] (TopMid5)  -- (Bouter);
\draw[proj] (LeftMid4) -- (Jouter);
\draw[proj] (LeftMid6) -- (Louter);

\end{tikzpicture}
}

\caption{
An $N$-colored Young diagram
$\lambda=(N+1,N,N-1,\ldots,3,2,1)$ and its corresponding Maya
diagram. Here $``\ldots"$ in $\lambda$ can be chosen arbitrarily,
provided that the entries satisfy the defining condition of a Young
diagram. Black dots and white dots of the Maya diagram indicate occupied  and empty sites respectively. Likewise, gray dots indicate intermediate Maya sites whose
occupancies are not shown explicitly.  All $\psi_i^{(j)}$s denote the vacuum Maya
diagram with color label.
}
\label{fig:young-maya-intersections}

\endgroup
\end{figure}
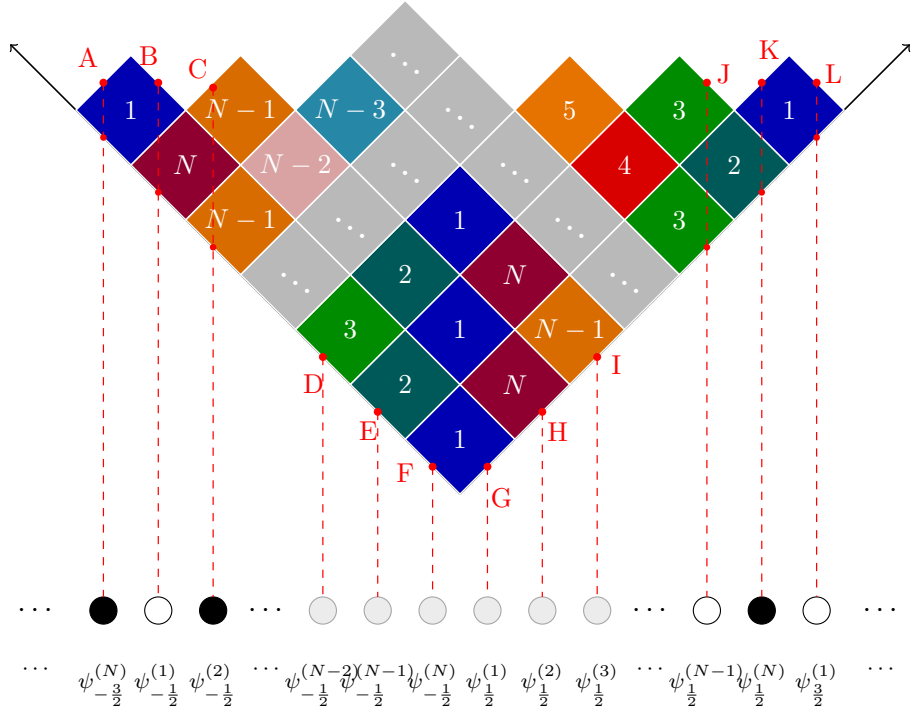

To get a better understanding of the convention mentioned above, we consider an example of an $N$-colored Young diagram and its corresponding Maya diagram in Figure \ref{fig:young-maya-intersections}.
For instance, the first row of the present Young diagram contains $N+1$ boxes, which represents the excited state $\psi_{-\frac{3}{2}}^{(N)}\psi_{-\left(\frac{1}{2}\right)}^{*(1)}|0\rangle$. Similarly, the second row of the Young diagram contributes a pair of fermionic operators $\psi_{-\frac{1}{2}}^{(2)}\psi_{-\left(\frac{1}{2}\right)}^{*(2)}$. Likewise each row contributes a pair of fermionic operators.
Putting them together we obtain an excited state $\psi_{\frac{1}{2}}^{(N)}\psi_{-\left(\frac{3}{2}\right)}^{*(1)}\ldots
\psi_{-\frac{1}{2}}^{(4)}\psi_{-\left(\frac{1}{2}\right)}^{*(3)}
\psi_{-\frac{1}{2}}^{(2)}\psi_{-\left(\frac{1}{2}\right)}^{*(2)}
\psi_{-\frac{3}{2}}^{(N)}\psi_{-\left(\frac{1}{2}\right)}^{*(1)}
|0\rangle$.

Each row of the Young diagram corresponds to one displacement of an Maya particle, which simultaneously determines the two components $\underline{k_i}$ and $\overline{k_i}$ in the decomposition of the final occupied state
\[
k_i=\underline{k_i}-N\overline{k_i}.
\]
For convenience, we take the wedge labels for the vacuum state to be
$k_i^{\mathrm{vac}}=N+1-i$.
The wedge labels for the Maya diagram $\lambda$ are then written as
$k_i=\lambda_i+k_i^{\mathrm{vac}}=\underline{k_{i}}-N\overline{k_{i}}$.

Now let us explain the correspondence mentioned above.
Suppose that the $i$-th row of the Young diagram contributes the pair of the fermionic operators
\begin{equation*}
\psi_{r_i^{\mathrm{occ}}}^{(a_i^{\mathrm{occ}})}
\psi_{-r_i^{\mathrm{vac}}}^{\ast(a_i^{\mathrm{vac}})}
|0\rangle ,
\end{equation*}
then the difference between the subscripts determines the orbital labels
\begin{equation*}
\overline{k_i}
=
\overline{k_i^{\mathrm{vac}}}+r_i^{\mathrm{occ}}-r_i^{\mathrm{vac}}.
\end{equation*}
Likewise the superscripts determine the color labels in the following way
\begin{equation*}
\underline{k_i}
=
\bigl[
\underline{k_i^{\mathrm{vac}}}
+a_i^{\mathrm{vac}}
-a_i^{\mathrm{occ}}
\bigr]_N,
\end{equation*}
where $[m]_N$ denotes the representative of $m$ modulo $N$ in the set
${1,\ldots,N}$.

Using the same Young diagram $\lambda=(N+1,N,N-1,\ldots,3,2,1)$ introduced above, we show that how the Maya points are related to both the spin component $\underline{k_{i}}$ and the orbital component $\overline{k_{i}}$ in  Figure \ref{fig:table-form-N-colored-maya-reduced}.

\begin{figure}[H]
\centering
\begingroup
\resizebox{0.5\textwidth}{!}{%
\begin{tikzpicture}[
    x=1cm,
    y=0.72cm,
    font=\normalsize
]

\def\xk{0}
\def\xku{2.2}
\def\xkb{4.4}

\newcommand{\MayaStart}[1]{\def\MayaY{#1}}

\newcommand{\MayaDec}{%
  \pgfmathtruncatemacro{\MayaY}{\MayaY-1}%
}

\newcommand{\MayaRow}[5][]{%
  \node[#1] at (\xk,\MayaY)  {\(\displaystyle #3\)};
  \node[#1] at (\xku,\MayaY) {\(\displaystyle #4\)};
  \node[#1] at (\xkb,\MayaY) {\(\displaystyle #5\)};

  \coordinate (#2Left)  at ({\xk-0.35},\MayaY);
  \coordinate (#2Right) at ({\xk+0.35},\MayaY);

  \MayaDec
}

\newcommand{\MayaVacuumLine}[1]{%
  \draw[dashed, thick] (-0.7,\MayaY) -- (5.4,\MayaY);
  \node[anchor=west] at (5.55,{\MayaY+0.2}) {#1};
  \MayaDec
}


\newcommand{\MayaArrowLeft}[6][]{%
  \draw[->, thick, #1]
    (#2Left)
    .. controls ($(#2Left)+(-#4,0)$) and ($(#3Left)+(-#4,0)$)
    ..
    node[
      pos=#5,
      above=2pt,
      xshift=-2pt,
      fill=white,
      inner sep=1pt
    ]
    {\(\displaystyle #6\)}
    (#3Left);
}

\newcommand{\MayaArrowRight}[6][]{%
  \draw[->, thick, #1]
    (#2Right)
    .. controls ($(#2Right)+(#4,0)$) and ($(#3Right)+(#4,0)$)
    ..
    node[
      pos=#5,
      right=2pt,
      yshift=1pt,
      fill=white,
      inner sep=1pt
    ]
    {\(\displaystyle #6\)}
    (#3Right);
}

\MayaStart{12}

\node at (\xk,{\MayaY+1.2})  {\(\displaystyle k\)};
\node at (\xku,{\MayaY+1.2}) {\(\displaystyle \underline{k}\)};
\node at (\xkb,{\MayaY+1.2}) {\(\displaystyle \overline{k}\)};

\MayaRow[blue]{topBlue}{2N+1}{1}{-2}
\MayaRow{topTwoN}{2N}{N}{-1}
\MayaRow[orange!85!black]{topRed}{2N-1}{N-1}{-1}
\MayaRow{topNmTwo}{2N-2}{N-2}{-1}
\MayaRow[cyan!60!black]{topOrange}{2N-3}{N-3}{-1}
\MayaRow{dotsUp}{\vdots}{\vdots}{\vdots}
\MayaRow{NpTwo}{N+2}{2}{-1}
\MayaRow{NpOne}{N+1}{1}{-1}

\MayaVacuumLine{ Dirac sea surface}

\MayaRow{vacN}{N}{N}{0}
\MayaRow{vacNmOne}{N-1}{N-1}{0}
\MayaRow{vacNmTwo}{N-2}{N-2}{0}
\MayaRow{dotsMid}{\vdots}{\vdots}{0}
\MayaRow[red]{one}{1}{1}{0}
\MayaRow{zero}{0}{N}{1}
\MayaRow{negOne}{-1}{N-1}{1}
\MayaRow{negTwo}{-2}{N-2}{1}
\MayaRow{dotsTail}{\vdots}{\vdots}{\vdots}


\MayaArrowLeft[blue]{vacN}{topBlue}{1.95}{0.60}{+\lambda_1}

\MayaArrowLeft[orange!85!black]{vacNmOne}{topRed}{1.45}{0.58}{+\lambda_2}

\MayaArrowLeft[cyan!60!black]{vacNmTwo}{topOrange}{1.05}{0.56}{+\lambda_3}

\MayaArrowRight[blue!70!black]{zero}{one}{0.45}{0.50}{+\lambda_{N+1}}

\end{tikzpicture}
}

\endgroup

\caption{For $\lambda=(N+1,N,N-1,\ldots,3,2,1)$, these wedge labels $k_i$, color labels $\underline{k_i}$, and orbital labels $\overline{k_i}$ are denoted as those colored rows. $k_i$s are obtained from the vacuum configuration by shifting $\lambda_i$ steps. For example, $N,N-1,N-2$ move to $2N+1,2N-1,2N-3$, respectively, while the last step shows that one particle moves from $0$ to $1$.  }
\label{fig:table-form-N-colored-maya-reduced}

\end{figure}

We still take Figure \ref{fig:young-maya-intersections} as an example. For the first row of the Young diagram, the corresponding pair of fermionic operators is written as
\begin{equation*}
\psi_{-\frac{3}{2}}^{(N)}
\psi_{-\left(\frac{1}{2}\right)}^{\ast(1)}
|0\rangle .
\end{equation*}
$\psi_{-\frac{1}{2}}^{\ast(1)}$ denotes denotes the annihilation of a fermion at $\frac{1}{2}^{(1)}$, while $\psi_{-\frac{3}{2}}^{(N)}$ denotes a fermion is created at $-\frac{3}{2}^{(N)}$.
The difference between these two sites is
\begin{equation*}
-\frac{3}{2}-\frac{1}{2}=-2,
\end{equation*}
which gives
\begin{equation*}
\overline{k_1}=-2.
\end{equation*}
On the other hand, the color changes from $1$ to $N$. Since the first vacuum spin label is
\begin{equation*}
\underline{k_1^{\mathrm{vac}}}=N,
\end{equation*}
the final spin label is obtained by adding the color difference $1-N$ to $\underline{k_1^{\mathrm{vac}}}$:
\begin{equation*}
\underline{k_1}=N+(1-N)=1.
\end{equation*}
Thus the wedge label is
\begin{equation*}
k_1=\underline{k_1}-N\overline{k_1}
=1-N(-2)=2N+1.
\end{equation*}

Similarly, for the second row, the corresponding pair of fermionic operator is
\begin{equation*}
\psi_{-\frac{1}{2}}^{(2)}
\psi_{-\left(\frac{1}{2}\right)}^{\ast(2)}
|0\rangle .
\end{equation*}
The  difference between these two sites is
\begin{equation*}
-\frac{1}{2}-\frac{1}{2}=-1,
\end{equation*}
which gives
\begin{equation*}
\overline{k_2}=-1.
\end{equation*}
The color does not change, since both color labels are $2$. The second vacuum spin label is 
\begin{equation*}
\underline{k_2^{\mathrm{vac}}}=N-1,
\end{equation*}
for which we have
\begin{equation*}
\underline{k_2}=(N-1)+(2-2)=N-1.
\end{equation*}
Therefore we obtain
\begin{equation*}
k_2=\underline{k_2}-N\overline{k_2}
=(N-1)-N(-1)=2N-1.
\end{equation*}

For the third row, the corresponding pair of fermionic operator is
\begin{equation*}
\psi_{-\frac{1}{2}}^{(4)}
\psi_{-\left(\frac{1}{2}\right)}^{\ast(3)}
|0\rangle .
\end{equation*}
The  difference between the sites is
\begin{equation*}
-\frac{1}{2}-\frac{1}{2}=-1,
\end{equation*}
and hence
\begin{equation*}
\overline{k_3}=-1.
\end{equation*}
The color changes from $3$ to $4$. Since the third vacuum spin label
\begin{equation*}
\underline{k_3^{\mathrm{vac}}}=N-2,
\end{equation*}
the final spin label is
\begin{equation*}
\underline{k_3}=(N-2)+(3-4)=N-3.
\end{equation*}
Then we have
\begin{equation*}
k_3=\underline{k_3}-N\overline{k_3}
=(N-3)-N(-1)=2N-3.
\end{equation*}
These wedge labels $k_i$, color labels $\underline{k_i}$, and orbital labels $\overline{k_i}$ have been shown in Figure \ref{fig:table-form-N-colored-maya-reduced}.

Each row of the Young diagram corresponds to one Maya displacement, and this displacement simultaneously determines the spin component $\underline{k_i}$ and the orbital component $\overline{k_i}$ of the final occupied state. For \(N=2\), the resulting two-colored Maya/Young-diagram
correspondence agrees with the chessboard example described in
Ref.~\cite{Chistyakova:2021yyd}.

 \section{Conclusions}\label{sec:summary}

In this paper, we have compared the hierarchy
generated by trigonometric Cherednik-Dunkl operators with that generated
by rational Dunkl operators. On the antisymmetric coordinate-spin space, the coordinate exchange operators are converted into
spin exchange operators through the relation $K_{ij}=-P_{ij}$.
Thus the two Dunkl-type hierarchies can be tdentified with the higher
Hamiltonians of the trigonometric and rational sCS models.

We have established two nested-commutator relations
\eqref{eq:tri_to_rat} and \eqref{rat to tri}. The  relation \eqref{eq:tri_to_rat} provides
an exact reconstruction of the rational hierarchy from the trigonometric
hierarchy by applying
the $m$-fold adjoint action of $E_{-1}$ to  $H_m^{\rm tri}$. In contrast, the relation \eqref{rat to tri} reconstructs the higher trigonometric sCS
Hamiltonians as leading terms from the rational ones by applying the
$m$-fold adjoint action of $E_1$ to  $H_m^{\rm rat}$.  

We have also explained the correspondence between $N$-colored Young
diagrams and Maya diagrams. In this correspondence, each part
$\lambda_i$ of the Young diagram records the displacement of the
$i$-th fermion from the Dirac sea in the vacuum Maya diagram. We found that the orbital component $\overline{k_i}$ is determined by the displacement of the Maya position, while the change of
the color label determines the spin component $\underline{k_i}$ together
with the vacuum spin label $\underline{k_i^{\mathrm{vac}}}$. Therefore,
the final occupied position
$
k_i=\lambda_i+k_i^{\mathrm{vac}}
$
is equivalently described by
$
k_i=\underline{k_i}-N\overline{k_i}.
$
This correspondence provides a direct combinatorial description of the spin,
momentum, and energy eigenvalue data of the trigonometric sCS Hamiltonian
hierarchy.

A natural problem for future work is to determine whether the lower
Dunkl-order terms appearing in the nested structure from the rational
hierarchy to the trigonometric hierarchy can be organized recursively in
terms of lower Hamiltonians of the sCS models.

Another interesting direction concerns supersymmetric CS models. Since
their Hamiltonians are formulated in terms of supersymmetric Dunkl-type
operators \cite{Desrosiers:2001ri}, it would be interesting to investigate
whether an analogous nested-commutator structure exists between the higher
Hamiltonians of supersymmetric trigonometric and rational CS hierarchies.

\section*{Acknowledgments}
We are grateful to Prof. Wei-Zhong Zhao for his helpful discussions.
This work is supported by the National Natural Science Foundation of China (No. 12375004).

\appendix
\section{The power sum description of the trigonometric Hamiltonians}
For completeness, we briefly mention the key steps leading to
\eqref{eq:powersum_H2} and \eqref{eq:powersum_H3}.
On a space of power sums
\[
p_m=\sum_{i=1}^{n}z_i^m,
\qquad
\partial_m:=\frac{\partial}{\partial p_m},
\]
the chain rule leads to
\begin{equation}\label{eq:power-chain}
\frac{\partial}{\partial z_i}
=
\sum_{a\geq1}a z_i^{a-1}\partial_a .
\end{equation}

For the second Hamiltonian, we rewrite the terms appearing in
\eqref{H2degenerate: rat to tri} in terms of power-sum variables.
The differential part is expressed by
\[
\sum_{i=1}^{n}(z_i\frac{\partial}{\partial z_i})^2
=
\sum_{a,b\geq1}
ab\,p_{a+b}\partial_a\partial_b
+
\sum_{a\geq1}a^2p_a\partial_a ,
\]
while the two-body contribution is reduced by the identity
\begin{equation}\label{eq:power-H2-key}
\sum_{i<j}
\frac{(z_i+z_j)(z_i^m-z_j^m)}
     {z_i-z_j}
=
\sum_{\substack{r+s=m\\r,s\geq1}}p_rp_s
+(n-m)p_m .
\end{equation}
Substituting these two expressions into
\eqref{H2degenerate: rat to tri}, we obtain the power-sum realization
\eqref{eq:powersum_H2}.

For the third Hamiltonian, the free third-order term becomes
\begin{equation}\label{eq:power-H3-free}
\begin{aligned}
\sum_i z_i^3(\frac{\partial}{\partial z_i})^3
={}&
\sum_{a,b,c\geq1}
abc\,p_{a+b+c}\partial_a\partial_b\partial_c
\\
&+
\frac32\sum_{a,b\geq1}
ab(a+b-2)p_{a+b}\partial_a\partial_b
\\
&+
\sum_{a\geq1}
a(a-1)(a-2)p_a\partial_a .
\end{aligned}
\end{equation}
The two-body sums appearing in
\eqref{degenerate: rat to tri} are reduced to
\begin{align}
\sum_{i\neq j}
\frac{z_i^m(z_i+z_j)}{z_i-z_j}
&=
\sum_{\substack{r+s=m\\r,s\geq1}}p_rp_s
+(n-m)p_m ,
\label{eq:power-Am}
\\
\sum_{i\neq j}
\frac{z_i^m(3z_i+2z_j)}{z_i-z_j}
&=
\frac52
\sum_{\substack{r+s=m\\r,s\geq1}}p_rp_s
+
\frac{6n-5m-1}{2}p_m .
\label{eq:power-Bm}
\end{align}
The only genuinely new ingredient is the three-body contribution.
Using the chain rule
(\ref{eq:power-chain}),
the three-body term in \eqref{degenerate: rat to tri} is transformed into
\[
\sum_{a>0}
\sum_{i\neq j, i\neq k, k\neq j}
\frac{
z_i(2z_i^2+3z_iz_j+3z_iz_k+2z_jz_k)
}{
(z_i-z_j)(z_i-z_k)
}
az_i^{a-1}\partial_a .
\]
To simplify the three-body contribution, we introduce
\[
G_m:=\sum_{i\neq j, i\neq k,k\neq j}
\frac{
z_i^m(2z_i^2+3z_iz_j+3z_iz_k+2z_jz_k)
}{
(z_i-z_j)(z_i-z_k)
}.
\]
We evaluate \(G_m\) by decomposing the sum into contributions from unordered triples of distinct variables. Thus, we first consider a fixed unordered triple
$
\{x,y,z\}.
$
For a fixed leading variable, the expression is symmetric in the other two variables. We use
\[
\sum_{\mathrm{cyc}}F(x,y,z)
:=F(x,y,z)+F(y,z,x)+F(z,x,y)
\]
for the sum over the three cyclic permutations of $(x,y,z)$. Since the summation over $j,k$ is ordered, the contribution of the unordered triple $\{x,y,z\}$ is
\[
2\sum_{\mathrm{cyc}}
\frac{x^m(2x^2+3xy+3xz+2yz)}{(x-y)(x-z)}.
\]
To rewrite the numerator in a form suitable for applying the Lagrange interpolation identity, we introduce the elementary symmetric polynomials
\[
e_1=x+y+z,\qquad e_2=xy+xz+yz.
\]
The numerator can then be expressed as
\[
2x^2+3xy+3xz+2yz=x^2+e_1x+2e_2.
\]
Therefore the fixed-triple contribution is
\begin{equation}\label{eq:Gm_xyz}
  2\left[
\sum_{\mathrm{cyc}}\frac{x^{m+2}}{(x-y)(x-z)}
+e_1\sum_{\mathrm{cyc}}\frac{x^{m+1}}{(x-y)(x-z)}
+2e_2\sum_{\mathrm{cyc}}\frac{x^m}{(x-y)(x-z)}
\right].
\end{equation}
We use the standard Lagrange interpolation identity
\[
\sum_{\mathrm{cyc}}
\frac{x^r}{(x-y)(x-z)}
=h_{r-2}(x,y,z),\qquad r\geq2,
\]
where $h_q$ is the complete homogeneous symmetric polynomial. We set $h_q=0$ for $q<0$. Hence the fixed-triple contribution \eqref{eq:Gm_xyz} is
\[
2\left(h_m+e_1h_{m-1}+2e_2h_{m-2}\right).
\]
 Now we retain the full monomial expansion of the complete homogeneous symmetric polynomial
\[
h_m(x,y,z)
=
\sum_{\substack{\alpha,\beta,\gamma\geq0\\
\alpha+\beta+\gamma=m}}
 x^\alpha y^\beta z^\gamma.
\]
For a monomial
\[
M_{\alpha,\beta,\gamma}=x^\alpha y^\beta z^\gamma,
\qquad
\alpha+\beta+\gamma=m,
\]
we denote by \(s=s(\alpha,\beta,\gamma)\) the number of nonzero exponents among
\(\alpha,\beta,\gamma\).

The monomial $M_{\alpha,\beta,\gamma}$ occurs exactly once in $h_m$. Since $e_1=x+y+z$, it can be obtained from $e_1h_{m-1}$ by selecting one of the $s$ variables that actually occurs in $M_{\alpha,\beta,\gamma}$. Hence
\[
[M_{\alpha,\beta,\gamma}](e_1h_{m-1})=s,
\]
where $[M_{\alpha,\beta,\gamma}]F(x,y,z)$ denotes the coefficient of the monomial $M_{\alpha,\beta,\gamma}=x^\alpha y^\beta z^\gamma$ in the polynomial $F(x,y,z)$.
Similarly, since $e_2=xy+xz+yz$, to obtain $M_{\alpha,\beta,\gamma}$ from $e_2h_{m-2}$, we choose a pair of distinct variables that both occur in the monomial. There are $\binom{s}{2}$ such pairs. With the pre-factor $2$ it gives rise to the coefficient
\[
 [M_{\alpha,\beta,\gamma}](2e_2h_{m-2})
=2\binom{s}{2}.
\]
Now we add the three contributions.
In particular, we obtain
\[
\begin{array}{c|c|c}
 s & \text{coefficient without the factor }2 & \text{final coefficient}\\ \hline
 1 & 1+1+0=2 & 4\\
 2 & 1+2+2=5 & 10\\
 3 & 1+3+6=10 & 20
\end{array}
\]
Thus the numbers $4$, $10$, and $20$ arise solely from the number of ways in which the variables occurring in a monomial which can be selected from $h_m$, $e_1h_{m-1}$, and $2e_2h_{m-2}$.

We define the auxiliary sums
$$
E_2(m):=
\sum_{i<j}
\sum_{\substack{r+s=m\\r,s\geq1}}
z_i^r z_j^s, \quad  E_3(m):=
\sum_{i<j<k}
\sum_{\substack{r+s+t=m\\r,s,t\geq1}}
z_i^r z_j^s z_k^t,
$$
which will be used below.
When \eqref{eq:Gm_xyz} is generalized to $n$ variables, a three-variable monomial contributes $20E_3(m)$. A two-variable monomial belongs to $n-2$ unordered triples, since the third variable can be chosen in $n-2$ ways. Hence a two-variable monomial  contributes  $10(n-2)E_2(m)$. Similarly a one-variable monomial $z_i^m$ belongs to $\binom{n-1}{2}$ unordered triples. Thus  a one-variable monomial contributes $4\binom{n-1}{2}p_m$. Therefore we derive
\begin{equation}\label{Gm}
G_m=20E_3(m)+10(n-2)E_2(m)+4\binom{n-1}{2}p_m.
\end{equation}
It remains to express $E_2(m)$ and $E_3(m)$ in terms of power sums.

For later use, we introduce the positive-index convolutions
\begin{equation}\label{Cm}
 C_m^{(2)}
:=
\sum_{\substack{r+s=m\\ r,s\geq 1}}p_rp_s,
\qquad
C_m^{(3)}
:=
\sum_{\substack{r+s+t=m\\ r,s,t\geq 1}}p_rp_sp_t .
\end{equation}
In $C_m^{(2)}$, the one-variable terms give rise to $(m-1)p_m$, while the two-variable terms give rise to $2E_2(m)$. Therefore we have
\begin{equation}\label{eq:E2}
    E_2(m)=\frac12\left(C_m^{(2)}-(m-1)p_m\right).
\end{equation}
In $C_m^{(3)}$, the one-variable terms give rise to $\binom{m-1}{2}p_m$. The two-variable terms  $z_i^k z_j^l, (i\neq j)$, with $k+l=m$,
give rise to $3(m-2)E_2(m)$. Likewise the three-variable terms give rise to $6E_3(m)$. Thus we derive 
\[
C_m^{(3)}
=
\binom{m-1}{2}p_m
+3(m-2)E_2(m)
+6E_3(m).
\]
Solving this identity for \(E_3(m)\), we have
\begin{equation}\label{eq:E3}
    E_3(m)
=
\frac16\left[
C_m^{(3)}-3(m-2)E_2(m)-\binom{m-1}{2}p_m
\right].
\end{equation} 
Substituting \eqref{eq:E2} and \eqref{eq:E3} into (\ref{Gm}), we obtain
\[
\begin{aligned}
G_m
={}&
\frac{10}{3}C_m^{(3)}
+5(n-m)C_m^{(2)}\\
&+\left(
\frac{10}{3}m^2-5mn+2n^2-n+\frac23
\right)p_m.
\end{aligned}
\]

\bibliographystyle{JHEP.bst}
\bibliography{biblio}

\end{document}